\documentclass[aps,prb,twocolumn,groupedaddress,showpacs,10pt]{revtex4-1}

\usepackage{amsmath}
\usepackage{amssymb}
\usepackage{bm}
\usepackage[utf8]{inputenc}
\usepackage{mathptmx}
\usepackage{graphicx}
\usepackage{multirow}
\usepackage[bookmarks=true,colorlinks=true,linkcolor=blue,citecolor=blue,urlcolor=blue]{hyperref}
\usepackage[figure]{hypcap}

\begin{document}


\title{Thickness--dependent magnetic structure of ultrathin Fe/Ir(001) films: 
from spin--spiral states towards ferromagnetic order}


\author{A. De\'ak}
\email{deak.andris@gmail.com}

\author{L. Szunyogh}
\affiliation{Department of Theoretical Physics, Budapest University of Technology and Economics, Budafoki \'ut 8., HU-1111 Budapest, Hungary}

\author{B. Ujfalussy}
\affiliation{Research Institute for Solid State Physics and Optics of the Hungarian Academy of Sciences, Konkoly-Thege M.\ \'ut 29-33., HU-1121 Budapest, Hungary}

\date{\today}

\begin{abstract}
We present a detailed study of the ground--state magnetic structure  
of ultrathin Fe films on the surface of fcc Ir(001). 
We use the spin-cluster expansion technique in combination with the relativistic disordered local moment
scheme to obtain parameters of spin models and then determine
the favored magnetic structure of the system by means of a mean field approach and 
atomistic spin dynamics simulations. For the case of a single monolayer of Fe we find that 
layer relaxations very strongly influence the ground-state spin configurations, whereas
Dzyaloshinskii--Moriya (DM) interactions and biquadratic couplings also have remarkable effects.
To characterize the latter effect we introduce and analyze spin collinearity maps of the 
system. 
While for two monolayers of Fe we find a single-$q$ spin spiral as ground state due to 
DM interactions, for the case of four monolayers the system shows a noncollinear spin structure with nonzero net magnetization. These findings are consistent with experimental measurements indicating 
ferromagnetic order in films of four monolayers and thicker.
\end{abstract}


\maketitle







\section{Introduction}\label{intro}
By the end of the 20th century, in particular, since the birth of spintronics, 
thin films and nanostructures have gained increasing importance in industrial applications.
The increasing demand for ultra-high density magnetic data storage devices had been one of the greatest driving forces of research and development involving nanostructures. With current hard disk technology approaching the superparamagnetic limit, the study of finite temperature magnetism in thin films and nanostructures 
is inevitable.
\emph{Ab initio} electronic structure methods give, in general, a very good description of the ground--state properties of solids. When trying to describe complex magnetic structures or finite temperature magnetism 
these methods are often used to generate parameters of spin models. 

It is by now widely accepted that relativistic corrections to the Heisenberg model, especially  
the Dzyaloshinskii--Moriya (DM) interaction, 
play an important role in determining the magnetic ground state of some systems.\cite{bode-2007,ferriani-2008,udvardi-2008,antal-2008} 
Moreover, higher order interaction terms (multiple spin interactions) also have to be considered in many cases \cite{antal-2007,hardrat-2009,lounis-2010} to give an accurate description of magnetism. Very recent studies indicate that such interactions may even lead to the formation of exotic states like magnetic skyrmion lattices.\cite{heinze-2011} 

In the present work we demonstrate the use of spin models from first principles for the description of magnetism,
in particular, determining the ground--state spin configuration of thin film systems. We use the spin-cluster expansion (SCE) combined with the relativistic disordered local moment (RDLM) theory to obtain model parameters. The SCE-RDLM method has just been successfully applied to the IrMn$_3$/Co(111) interface, a prototype for an exchange bias system, to calculate exchange interactions and magnetic anisotropies.\cite{szunyogh-2011}

Here we go beyond the tensorial Heisenberg model\cite{udvardi-2003} by including biquadratic couplings 
in the spin Hamiltonian
and present our results for thin films of Fe on the (001) surface of fcc Ir. 
Previous theoretical studies have found that in case of a single Fe overlayer the favored magnetic configuration depends very strongly on layer relaxations, leading to the formation of spin spiral states near the experimental geometry.\cite{kudrnovsky-2009}
We study the effect of layer relaxations on the magnetic interactions for the case of a monolayer, as well as the effect of higher order interactions on the spin configuration. We also examine thicker films consisting of two and four layers of Fe to see whether the bulk ferromagnetism of Fe emerges with increasing film thickness. 
Experiments have shown that thin films of Fe consisting of 4 or more monolayers produce a ferromagnetic signal in magneto-optic Kerr effect (MOKE) measurements with an in-plane easy axis.\cite{martin-2007} 
The results of our simulations turn to be consistent with the main experimental findings.

\section{Theory}\label{theory}
\subsection{Spin model}
The most widely used ab initio methods to describe itinerant magnetic systems rely
on the adiabatic approximation separating fast single--electron spin fluctuations from 
the slow transversal motion of the spins.\cite{antropov-1996}
Furthermore, in the so-called rigid spin approximation it is assumed that longitudinal fluctuations 
of the local moments are negligible, so that the system of $N$ moments is characterized by a set of unit vectors $\left\lbrace \vec e\right\rbrace=\left\lbrace \vec e_1, \dotsc, \vec e_N\right\rbrace$ describing the orientation of each local moment. Then the grand potential $\Omega\!\left(\left\lbrace \vec e\right\rbrace\right)$ of the system may be thought of as a classical spin Hamiltonian,\cite{gyorffy-1985} which can be used in numerical simulations to study the various magnetic properties of the system.

For practical applicability, the energy must be parametrized in a simple yet meaningful way. The most common approximation is a form of a generalized Heisenberg model,
\begin{align}
\Omega\!\left(\left\lbrace \vec e\right\rbrace\right)=\Omega_0 +\sum\limits_{i=1}^N \vec e_i \,\mathbf K_i \vec e_i -\frac{1}{2}\sum\limits_{\substack{i, j=1\\\left(i\ne j\right)}}^N \vec e_i \,\mathbf J_{ij}\vec e_j,\label{spinham-2ord}
\end{align}
where $\mathbf K_i$ and $\mathbf J_{ij}$ are the second order on-site anisotropy matrices and tensorial exchange couplings, respectively. The latter can be meaningfully decomposed into an isotropic component $J_{ij}^I=\operatorname{Tr}\mathbf J_{ij}/3$, an antisymmetric component $\mathbf J_{ij}^A$ and a traceless symmetric part $\mathbf J_{ij}^S$. The isotropic component describes a Heisenberg interaction, the antisymmetric part corresponds to the DM interaction\cite{dzyaloshinskii-1958,moriya-1960} in the form of
\begin{align}
\vec e_i \;\mathbf J_{ij}^A\, \vec e_j=\vec D_{ij} \left(\vec e_i \times \vec e_j\right),
\end{align}
and the final component contributes to the so-called two-site anisotropy. In this paper we go beyond the second order expansion of Eq.~\eqref{spinham-2ord} and include isotropic biquadratic interaction terms of the form $-B_{ij}\!\left(\vec e_i \cdot \vec e_j\right)^2$.

Even though a parametrization of the energy with a spin Hamiltonian possesses a much less direct connection to the magnetic properties of the system, it can be used well to simulate the magnetic behavior. One method for obtaining the parameters of the spin Hamiltonian directly from first principles calculations is the so-called spin-cluster expansion (SCE) combined with the relativistic disordered local moment scheme (RDLM).

\subsection{Spin-cluster expansion}
The spin-cluster expansion developed by Drautz and F\"ahnle\cite{drautz-2004,drautz-2005} gives a systematic parametrization of the adiabatic energy surface. Up to two-spin interactions, the grand potential may be expanded using real spherical harmonics as
\begin{align}
\Omega\!\left(\left\lbrace \vec e\right\rbrace\right) & \simeq \Omega_0 + \sum\limits_i \sum\limits_{L\ne \left(0,0\right)} J_i^L Y_L\!\left(\vec e_i\right)\notag\\
&\quad+\frac{1}{2}\sum\limits_{i\ne j} \sum\limits_{L\ne \left(0,0\right)} \sum\limits_{L^\prime \ne \left(0,0\right)} J_{ij}^{LL^\prime} Y_L\!\left(\vec e_i\right) Y_{L^\prime}\!\left(\vec e_j\right),\label{grandpot-sce}
\end{align}
where the summations do not include the constant spherical harmonic function of the composite index $\left(\ell,m\right)=\left(0,0\right)$. The coefficients in Eq.~\eqref{grandpot-sce} are defined as
\begin{align}
\Omega_0&=\left\langle \Omega \right\rangle,\label{sce-0site}\\
J_i^L&=\int \mathrm d^2 e_i \left\langle\Omega\right\rangle_{\!\vec e_i} Y_L\!\left(\vec e_i\right)\\
J_{ij}^{LL^\prime}&=\int \mathrm d^2 e_i \int \mathrm d^2 e_j \left\langle \Omega\right\rangle_{\!\vec e_i \,\vec e_j} Y_L\!\left(\vec e_i\right) Y_{L^\prime}\!\left(\vec e_j\right)\label{sce-2site},
\end{align}
where vectors in lower index indicate restricted averages, i.e., uniform directional averaging has to be carried out with respect to every spin in the system not noted in the lower index.

The terms of the spin Hamiltonian can be directly related to the terms of the SCE, for instance the isotropic biquadratic couplings can be expressed as
\begin{align}
B_{ij}&=-\frac{3}{8\pi}\sum\limits_{m=-2}^2 J_{ij}^{(2,m)(2,m)}.
\label{bij}
\end{align}
Clearly the key quantities of the SCE are the restricted directional averages of the grand potential. 

\subsection{Relativistic disordered local moment theory}
The DLM scheme gives a description of a magnetic system in accordance with the adiabatic approximation. Its implementation within the Korringa-Kohn-Rostoker (KKR) theory was given by Gy\"orffy \emph{et al.},\cite{gyorffy-1985} with a relativistic generalization by Staunton \emph{et al.}\cite{staunton-2004,staunton-2006} Combining it with the SCE provides a highly effective tool for determining the parameters of spin models. For a detailed presentation of the SCE-RDLM method see Ref.~\onlinecite{szunyogh-2011}. In the following we will review the most important features of the theory.

The electronic charge and magnetization densities are determined from a self-consistent-field KKR calculation. In good moment systems the magnitude of local moments may be considered as independent from their orientation. For a given set of self-consistent potentials, charge and local moment magnitudes, the orientations $\left\lbrace \vec e \right\rbrace$ of the local moments are accounted for by the similarity transformation of the single-site $t$-matrices,
\begin{align}
\underline t_i\!\left(\vec e_i\right)=\underline R\!\left(\vec e_i\right) \underline t_i\!\left(\vec e_z\right) \underline R\!\left(\vec e_i\right)^\dagger,
\end{align}
where $\underline t_i\!\left(\vec e_z\right)\equiv \underline t_i\!\left(\varepsilon;\vec e_z\right)$ is the $t$-matrix for a given energy, $\varepsilon$ (not labeled explicitly), 
with exchange field along the $z$ axis, and $\underline R\!\left(\vec e_i\right)$ is the representation of the $SO\!\left(3\right)$ rotation that transforms $\vec e_z$ into $\vec e_i$. Underlines denote matrices in the $\left(\kappa,\mu\right)$ angular momentum representation.

The coherent potential approximation (CPA) is employed to describe the magnetically disordered system. The strategy of the CPA is to substitute the disordered system with an effective (coherent) medium which is independent from the orientation of local moments, such that the scattering of an electron in the effective medium should resemble the average scattering in the disordered physical system. The scattering path operator of the effective medium is defined as
\begin{align}
\underline{\underline \tau}_c=\left(\underline{\underline t}_c^{-1}-\underline{\underline G}_0\right)^{-1},
\end{align}
where double underlines denote matrices in site-angular momentum space, $\underline{\underline G}_0$ is the matrix of structure constants, and $\underline{\underline t}_c$ is site diagonal. Using the excess scattering matrices defined as
\begin{align}
\underline X_i\!\left(\vec e_i\right)=\left[\left(\underline t_{c,i}^{-1}-\underline t_{i}^{-1}\!\left(\vec e_i\right)\right)^{-1} - \underline \tau_{c,ii}\right]^{-1}
\end{align}
the single-site CPA condition can be formulated as
\begin{align}
\int \mathrm d^2 e_i\, \underline X_i\!\left(\vec e_i\right)=\underline 0.
\end{align}

Within validity of the magnetic force theorem the grand potential of the system can be expressed in terms of the excess scattering matrices and the related impurity matrices
\begin{align}
\underline D_i\!\left(\vec e_i\right)=&\left[\underline I + \left(\underline t_i^{-1}\!\left(\vec e_i\right)-\underline t_{c,i}^{-1}\right) \underline \tau_{c,ii} \right]^{-1}
\end{align}
for a given spin configuration and at zero temperature as
\begin{align}
\Omega\!\left(\left\lbrace \vec e \right\rbrace\right)&=\Omega_c -\frac{1}{\pi}\sum\limits_i \operatorname{Im}\int\limits^{\varepsilon_F} \mathrm d\varepsilon \,\ln\det \underline D_i\!\left(\vec e_i\right)\notag\\
&\quad-\frac{1}{\pi}\sum\limits_{k=1}^\infty \frac{1}{k} \sum\limits_{i_1\ne i_2\ne \dotsb \ne i_k \ne i_1} \operatorname{Im}\int\limits^{\varepsilon_F} \mathrm d\varepsilon \operatorname{Tr}\big[\underline X_{i_1}\!\left(\vec e_{i_1}\right) \underline \tau_{c,i_1 i_2} \notag\\
&\qquad\times\underline X_{i_2}\!\left(\vec e_{i_2}\right) \dotsm \underline X_{i_k}\!\left(\vec e_{i_k}\right) \underline \tau_{c,i_k i_1}\big],\label{grandpot-rdlm}
\end{align}
where the constant term reads as
\begin{align}
\Omega_c=-\int\limits^{\varepsilon_F} \mathrm d\varepsilon \,N_0\!\left(\varepsilon\right)-\frac{1}{\pi}\operatorname{Im}\int\limits^{\varepsilon_F} \mathrm d\varepsilon \,\ln\det \underline{\underline \tau}_c\!\left(\varepsilon\right),
\end{align}
$N_0\!\left(\varepsilon\right)$ being the integrated density of states of free electrons. Eq.~\eqref{grandpot-rdlm} can be used to calculate restricted averages of the grand potential for the SCE, Eqs.~\eqref{sce-0site}--\eqref{sce-2site}. In particular the two-site expansion terms are 
expressed as
\begin{align}
J_{ij}^{LL^\prime}&=-\frac{1}{\pi}\operatorname{Im}\int \limits^{\varepsilon_F} \mathrm d\varepsilon  \iint \mathrm d^2 e_i \;\mathrm d^2 e_j\; Y_L\!\left(\vec e_i\right) Y_{L^\prime}\!\left(\vec e_j\right)\notag\\
&\qquad\times\operatorname{Tr}\ln\left[\underline I-\underline X_i\!\left(\vec e_i\right) \underline \tau_{c,ij}\underline X_j\!\left(\vec e_j\right)\underline \tau_{c,ji}\right],\label{sce-rdlm-2site}
\end{align}
implying that higher order two-site terms, such as biquadratic couplings, see Eq.~(\ref{bij}), 
can be calculated just as easily as tensorial Heisenberg interactions.

\section{Results}\label{results}
\subsection{Computational details}
We used the screened Korringa--Kohn--Rostoker (SKKR) method\cite{szunyogh-1994,zeller-1995} to perform self-consistent calculations for fcc bulk Ir and the layered Fe/Ir(001) systems. 
The in-plane lattice constant for the fcc lattice was chosen 2.715~\AA.\cite{martin-2007} We used the local spin-density approximation parametrized according to Ceperley and Alder,\cite{ceperley-1980} and we employed the atomic sphere approximation with an angular momentum cutoff of $\ell_\text{max}=3$. A scalar-relativistic DLM description was used, corresponding to the paramagnetic state. Matrices of the effective CPA medium were determined to a relative error of 10$^{-5}$. We used 12 energy points on a semicircular path on the upper complex half-plane for the energy integrations, and 78 points were sampled in the irreducible wedge of the 2D Brillouin zone (BZ) for $k$-integrations.

The spherical integrations needed for the SCE, as in Eq.~\eqref{sce-rdlm-2site}, were performed according to the Lebedev--Laikov scheme.\cite{lebedev-1999} The logarithm in Eq.~\eqref{sce-rdlm-2site} was expanded into a power series to avoid the phase problem due to the energy integration of the complex logarithm.

\subsection{Fe$_1$/Ir(001)}
\begin{figure}
\includegraphics[width=\linewidth]{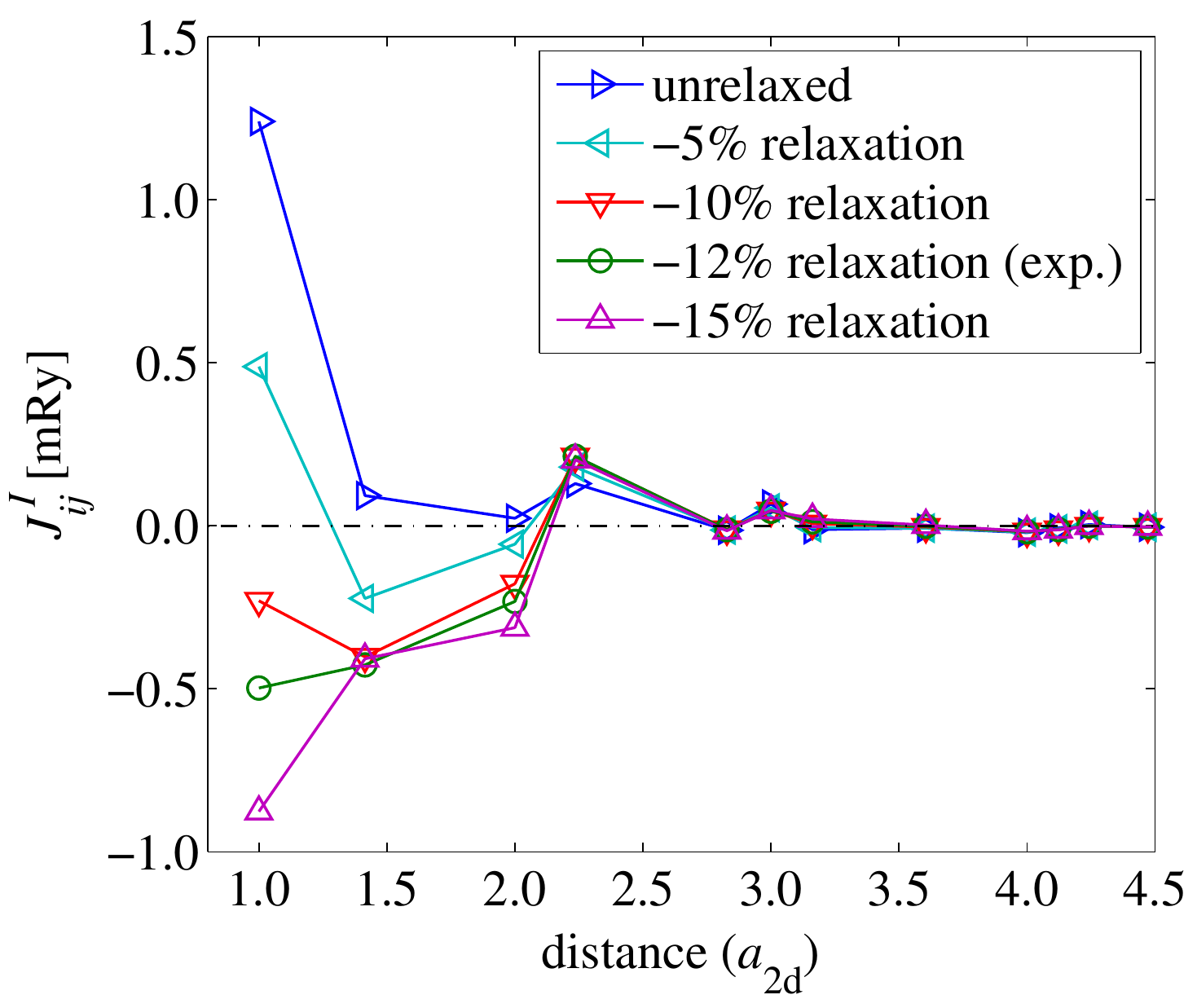}
\caption{(Color online) Isotropic couplings for various layer relaxations in $\text{Fe}_1/\text{Ir}(001)$ as a function of interatomic distance in units of the in-plane lattice constant, $a_\text{2d}$. }
\label{fe1ir-isos}
\end{figure}

Firstly we performed calculations for the case of an Fe monolayer. Along with
the unphysical case of an unrelaxed Fe monolayer, systems with different layer
relaxations were considered, namely 5\%, 10\%, 15\% inward relaxations as well
as the experimental value of -12\% relaxation.\cite{martin-2007} According to
Kudrnovsk\'y \emph{et al.},\cite{kudrnovsky-2009} we may expect a very strong
dependence of exchange parameters on the layer relaxation, with a crossover of
dominant Heisenberg couplings from ferromagnetic (FM) to antiferromagnetic
(AFM). The obtained isotropic couplings are shown in Fig.~\ref{fe1ir-isos}.

While for the unrelaxed geometry all significant couplings are FM, with increasing inward relaxation the couplings for the three nearest neighbor (NN) shells become gradually AFM. For the case of experimental layer relaxation, the magnitude of the Heisenberg interactions is small and comparable up to four shells. While the tendency of the obtained curves is similar to those calculated by Kudrnovsk\'y \emph{et al.},\cite{kudrnovsky-2009}  there are remarkable differences. In particular, at the experimental layer relaxation our largest isotropic couplings are 1~mRy smaller than those obtained in Ref.~\onlinecite{kudrnovsky-2009}, corresponding to a difference by a factor of three. Since the TB-LMTO-SGF method is known to give results that agree very well with KKR we use, we performed a series of check calculations to verify the reliability of our results. Increasing the numerical precision of the spherical integrations of the SCE, the convergence parameters of the CPA and the number of $k$ points for the BZ integration all resulted in identical isotropic couplings within linewidth. Increasing the number
of energy points from 12 to 16 used for the energy integrations in the self-consistent calculations 
also leads to relative differences less then 2\%. 

We also performed calculations with spin-orbit coupling scaled to zero\cite{ebert-1996} as well as with the parametrization of the LDA according to Vosko, Wilk and Nusair.\cite{vosko-1980} The differences in the dominating exchange interactions from both sets of calculations were less then 0.1~mRy indicating that such factors are insufficient to explain the quantitative difference between our interactions and those of Kudrnovsk\'y \emph{et al}.\cite{kudrnovsky-2009} 

Due to the frustrated nature and small magnitude of the isotropic exchange interactions it is possible that
 DM interactions or biquadratic couplings affect the ground--state spin configuration. The magnitude of these interactions for the first two NN\ shells is shown in Table~\ref{fe1ir_dms_bijs}. For higher relaxation values, in particular for the experimental one, the correction terms are indeed comparable to the isotropic couplings. 
Interestingly, in our case the biquadratic couplings all have positive sign indicating that these interactions favor collinear spin configurations. This feature implies a competition between the frustrated AFM exchange couplings, the DM and the biquadratic interactions, possibly leading to complex ground--state spin configurations.
\begin{table}
\caption{Calculated DM interactions and isotropic biquadratic couplings, see Eq.~(\ref{bij}), in $\text{Fe}_1/\text{Ir}(001)$. All values are given in mRy units.}
\label{fe1ir_dms_bijs}
\begin{ruledtabular}
\begin{tabular}{ccccccc}
\multicolumn{2}{r}{Layer relaxation}    &  0\% & -5\% & -10\% & -12\% & -15\%\\
\colrule
\multirow{2}{*}{$\left\lvert \vec D_{ij}\right\rvert$} & 1st shell  & 0.023 & 0.142 & 0.244 & 0.280 & 0.331 \\
                                                              & 2nd shell & 0.156 & 0.207 & 0.209 & 0.198 & 0.170 \\
\multirow{2}{*}{$B_{ij}$} & 1st shell  & 0.110 & 0.109 & 0.088 & 0.077 & 0.059 \\
                                & 2nd shell & 0.008 & 0.012 & 0.013 & 0.013 & 0.012\\
\end{tabular}
\end{ruledtabular}
\end{table}

The Fourier transform of the tensorial coupling matrices was calculated to obtain the mean field estimate, as explained in the \hyperref[app:spinsusc]{Appendix}. In the monolayer case the Fourier transform is a $3\times 3$ tensor field defined on the two-dimensional BZ. The largest eigenvalue $J\!\left(\vec q\right)$ can be visualized as a surface in reciprocal space. The mean field estimate states that the maximum points of the surface give the characteristic spatial modulation of the ordered magnetic state, to which the paramagnetic state is unstable.

\begin{figure}
\includegraphics[width=\linewidth]{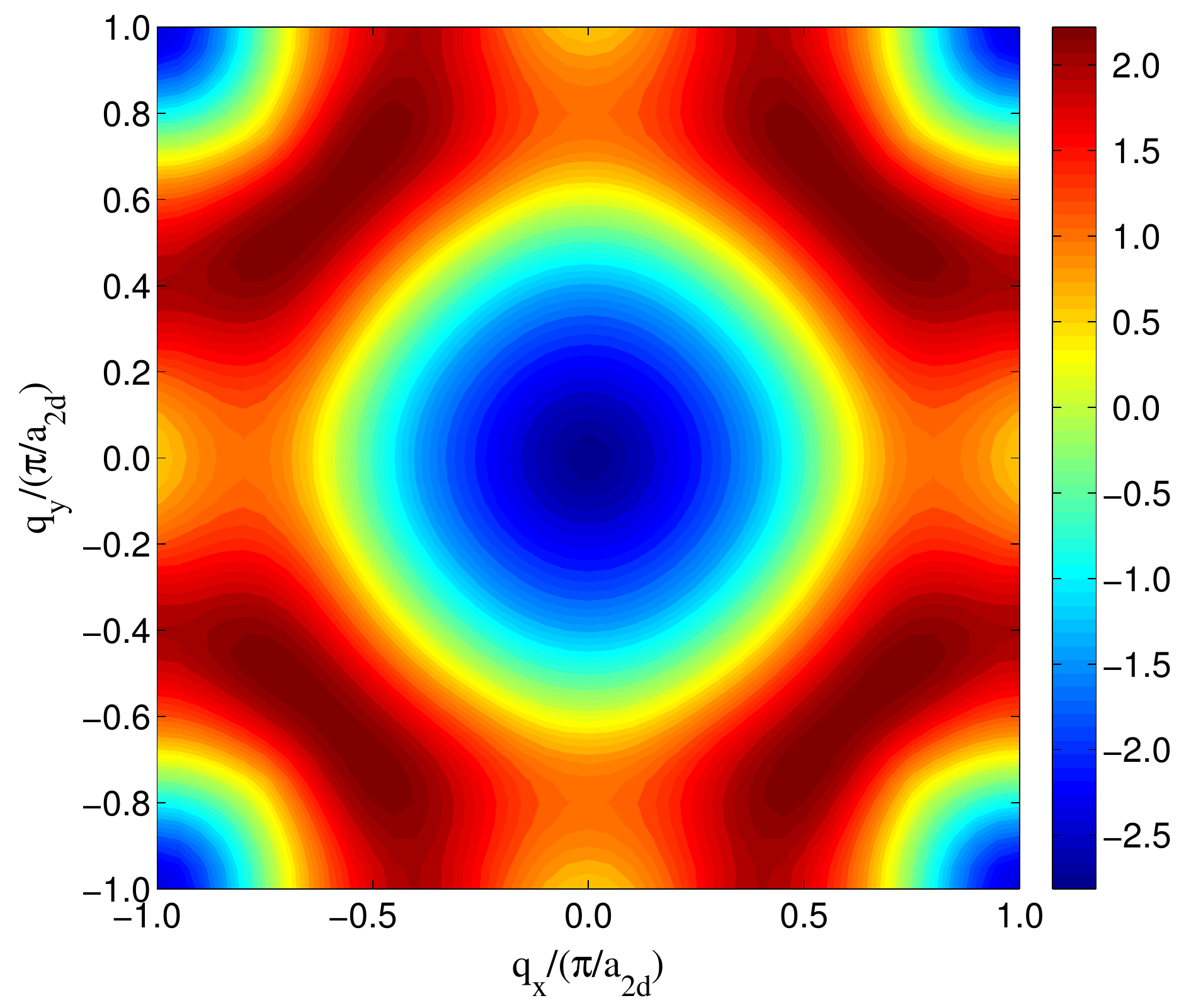}
\caption{(Color online) $J\!\left(\protect\vec q\right)$ surface for Fe$_1$/Ir(001) with experimental relaxation.
The color coding in units of mRy is shown on the right.}
\label{fe1ir_r12_dlm_jq}
\end{figure}

Reassuringly, the $J\!\left(\vec q\right)$ surface for the unrelaxed system has a pronounced maximum in the $\Gamma$ point of the BZ, indicating FM ground state. As the layer relaxation is increased, the corresponding surfaces are gradually depressed at the $\Gamma$ point, giving way to new maxima in general points of the 
Brillouin zone for -10\% and larger inward relaxations. The contour plot of the $J\!\left(\vec q\right)$ surface for experimental layer relaxation is shown in Fig.~\ref{fe1ir_r12_dlm_jq}. The maxima are around $\vec q=\left(0.5, 0.7\right)\!\frac{\pi}{a_\text{2d}}$ and symmetry related points. This estimate suggests a complex noncollinear spin structure due to the frustrated and competing magnetic interactions detailed earlier. Note that there is an approximate degeneracy along lines connecting pairs of maxima. The difference between the maximum values of the surface and its value at, for instance, $\vec q=\left(0.6, 0.6\right)\!\frac{\pi}{a_\text{2d}}$ is less then 0.03~mRy. This implies the possibility 
that the actual ground state of the system is somewhere along these degenerate lines, but not necessarily in the precise maxima of the surface.

The effect of relativistic tensorial interactions may be assessed by calculating the $J\!\left(\vec q\right)$ surface using only the isotropic part of the exchange tensors. In doing so, the shape of the resulting surface is the same, however its maxima are shifted closer to the zone boundary, to $\vec q=\left(0.45, 0.85\right)\!\frac{\pi}{a_\text{2d}}$ and related points. The degeneracy of the line connecting the maxima is 
remarkably decreased. 
Based on the mean field estimate, the ground--state spin configuration of Fe$_1$/Ir(001) is a spin spiral, 
the periodicity of which is largely affected by the DM interactions.

We performed zero temperature Landau--Lifshitz--Gilbert spin dynamics simulations to verify the mean field prediction and to reveal the influence of biquadratic couplings on this noncollinear spin structure. Due to the expected noncollinear structures we used a $128\times 128$ lattice of spins along with free boundary conditions to prevent the appearance of spurious periodicity. Interactions were included up to 7 two-dimensional lattice constants ($a_\text{2d}$). For every relaxation two kinds of simulations were performed, one with and the other without biquadratic coupling terms. Every simulation was initialized from a random spin configuration, the same for simulations with and without biquadratic terms. All simulations were continued until there was no measurable difference (i.e.,\ less than $10^{-6}$~mRy) in the total interaction energy of the system. 
It should be noted that on-site anisotropy was not included in the simulations due to its small size 
(0.06~mRy in case of experimental layer relaxation), however two-site anisotropy was included as the symmetric part of the $\underline{\underline J}_{ij}$ tensors.

In the case of unrelaxed geometry, the simulations led to (out-of-plane) FM order as expected. As layer relaxation is introduced, the transition predicted by the mean field estimates is reflected in the spin dynamics simulations as well, as the ground states become complex spin-spirals for relaxations of -~10\% and above. The Fourier components of the emergent spin configurations agree very well with the mean field estimates. Interestingly, for -15\% relaxation the spin configuration of the simulated system contains a huge number of domains with various orientations for the spin spirals, which is in very good agreement with the degeneracy of the $J\!\left(\vec q\right)$ surface along the corresponding $\vec q$ points. 

\begin{figure}
\includegraphics[width=\linewidth]{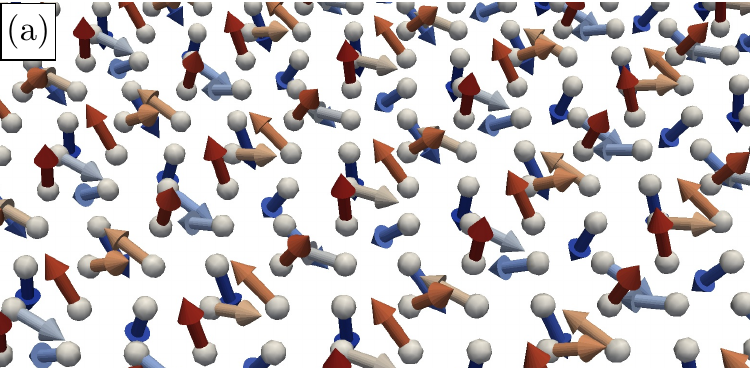}
\\
\includegraphics[width=\linewidth]{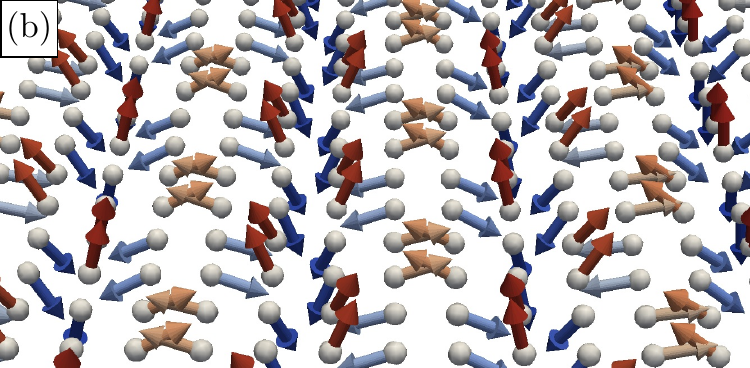}
\caption{(Color online) Ground--state spin configurations of the Fe monolayer with experimental layer relaxation (a)~\emph{without} and (b)~\emph{with} biquadratic couplings.}
\label{fe1ir-sd}
\end{figure}

It is expected that biquadratic couplings can significantly affect the ground--state spin configuration 
when the tensorial Heisenberg interactions alone lead to the formation of noncollinear spin structures. 
This is the case for the experimental layer relaxation for which the resulting spin-configurations
of the two simulations are shown in Fig.~\ref{fe1ir-sd}. 
The lattice Fourier transform of the spin configuration in Fig.~\ref{fe1ir-sd}a has a peak at $\vec q\approx \left(0.719, 0.495\right)\!\frac{\pi}{a_\text{2d}}$, which perfectly coincides with the numerical maximum of the corresponding $J\!\left(\vec q\right)$ surface. Interestingly the peak for the simulation including biquadratic terms is at $\vec q\approx \left(0.735, 0.500\right)\!\frac{\pi}{a_\text{2d}}$, which is only very slightly different from the previous value. Even though the two spin structures in Fig.~\ref{fe1ir-sd} seem very different, their spatial modulation is almost identical. The main reason for this is that while in Fig.~\ref{fe1ir-sd}b there is a clear two~row-by-two~row periodicity along one direction, in Fig.~\ref{fe1ir-sd}a we can see a helical spiral along the same direction with $90^\circ$ angle between adjacent spins. Both orderings have a period of $4a_\text{2d}$.

\begin{figure}[th]
\includegraphics[width=\linewidth]{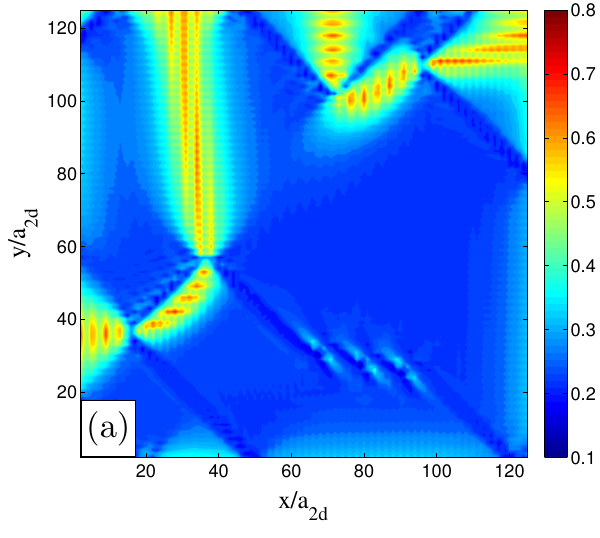}
\includegraphics[width=\linewidth]{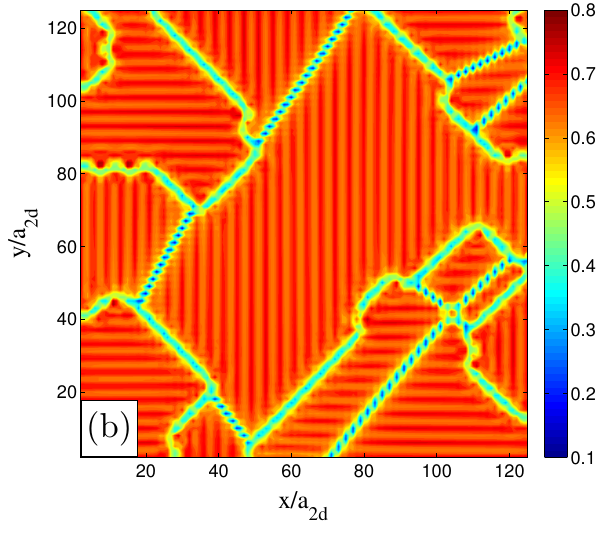}
\caption{(Color online) Collinearity map for the approximate ground state of Fe$_1$/Ir(001) with experimental layer relaxation, (a)~\emph{without} and (b)~\emph{with} biquadratic couplings.
The color coding in units of mRy is shown on the right.}
\label{fe1ir-coll}
\end{figure}

We may try to grasp the effect of biquadratic interactions based on the real space spin structure. From the form of the biquadratic coupling between two specific spins, $-B_{ij}\left(\vec e_i \cdot \vec e_j\right)^2$, it is clear that a coupling with positive sign favors collinear alignment, either parallel or antiparallel. It is motivated that positive biquadratic couplings acting on a noncollinear spin configuration will try to increase the collinearity within the system.

Let us define the collinearity of two spins as
\begin{align}
\operatorname{coll}\!\left(\vec e_i,\vec e_j\right)=\frac{2}{\pi} \left\lvert \arccos \left(\vec e_i\cdot\! \vec e_j\right)-\frac{\pi}{2}  \right\rvert,
\end{align}
which is simply the deviation of the angle between the two spins from a right angle, normalized between 0 and 1, 1 indicating most collinear arrangement (i.e.,\ where $\vec e_i\cdot\vec e_j=\pm 1$). Using this definition we may create a collinearity map of any spin configuration, plotting at each site an averaged collinearity defined as
\begin{align}
\operatorname{coll}\!\left(i\right)=\frac{1}{z_i}\sum\limits_{\substack{j\\ \left\langle i,j\right\rangle}}\operatorname{coll}\!\left(\vec e_i,\vec e_j\right),
\end{align}
where the summation includes every first NN\ of site $i$ and correspondingly $z_i$ stands for the coordination number of site $i$. The reason for the consideration of only first NN\ sites is, besides practicality, the 
rapid spatial decay of biquadratic couplings, suggesting that this interaction is mainly sensitive to 
the first NN\ collinearity.

Using the above definitions, the calculated collinearity maps from the two kinds of simulations are shown for experimental relaxation in Fig.~\ref{fe1ir-coll}. The difference between the two maps is remarkable. At first glance we can see that the two configurations have a completely different domain structure. While in Fig.~\ref{fe1ir-coll}a there are smoothly connected domains of low collinearity (0.22 on average), in Fig.~\ref{fe1ir-coll}b we can see neatly separated, homogeneous domains of high collinearity (0.67 on average). This is a clear indication that biquadratic couplings have a serious effect on the magnetic structure in this specific system.

\subsection{Fe$_\mathbf{2}$/Ir(001)}
Martin \emph{et al.}\cite{martin-2007} found using MOKE measurements that Fe films of 4 monolayers or thicker produce an in-plane FM signal at room temperature. For the case of an Fe monolayer with reasonable layer relaxations the theoretically predicted spin spiral states have zero net magnetization, thus they would not provide a MOKE signal. To see if the bulk-like behavior of Fe emerges for thicker films, we extended our studies to cases of two and four monolayers of Fe on Ir(001).

\begin{figure}
\includegraphics[width=\linewidth]{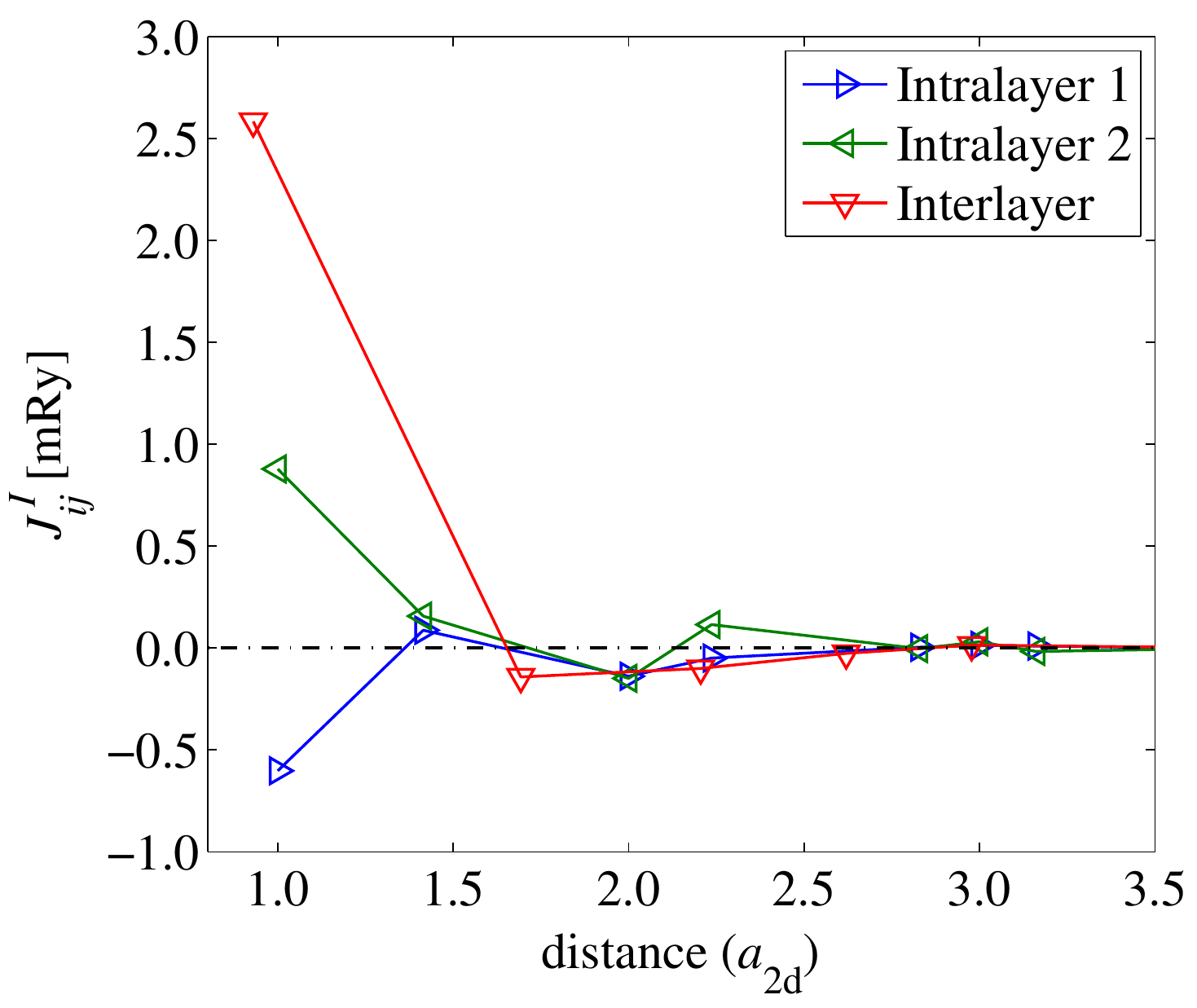}
\caption{(Color online) Isotropic Heisenberg couplings in $\text{Fe}_2/\text{Ir}(001)$ as a function of interatomic distance in units of the in-plane lattice constant, $a_\text{2d}$. }
\label{fe2ir-isos}
\end{figure}

The geometries of Fe$_n$/Ir(001), both for $n=2$ and $n=4$ were assumed according to Ref.~\onlinecite{martin-2007}, with the minor simplification that no layer relaxation was applied to the Ir substrate. As shown in Fig.~\ref{fe2ir-isos}, the isotropic Heisenberg couplings we obtained show a strong FM coupling between the two layers, while the intralayer couplings are smaller, and in case of the subsurface layer the dominant ones are AFM. It should be mentioned that the DM interactions are one order of magnitude smaller then the isotropic terms, and the biquadratic couplings are even smaller by one order.

The $J\!\left(\vec q\right)$ surface for the bilayer has its maximum near the $\Gamma$ point, but not exactly at the zone center. There is a degenerate circle at $\left\lvert \vec q\right\rvert\approx 0.07\frac{\pi}{a_\text{2d}}$ where the surface is maximal, predicting a long wavelength spin spiral. Considering only the isotropic part of the exchange tensors the corresponding surface has its maximum in the $\Gamma$ point. This indicates that the DM interaction imposes a noncollinear spin structure on the otherwise FM system.

\begin{figure}
\includegraphics[width=0.8\linewidth]{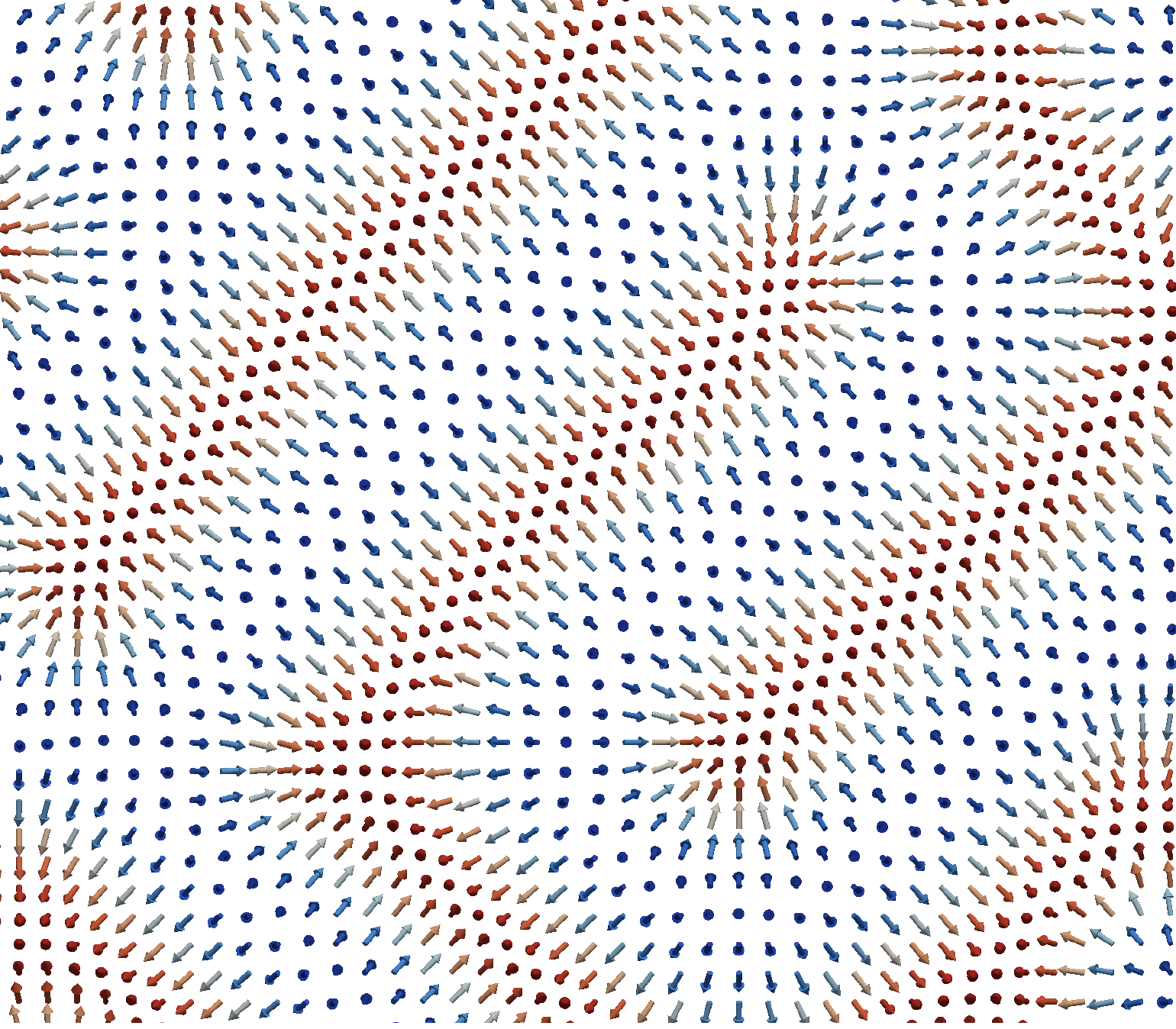}
\caption{(Color online) Approximate ground--state spin configuration of Fe$_2$/Ir(001), surface layer shown from the top. Arrows are colored according to $z$ component, ranging from red (up) to blue (down). }
\label{fe2ir-sd-withb}
\end{figure}

Since the number of neighbors within the same radius rapidly increases with the number of Fe layers, the spin dynamics simulations for the multilayers were performed with in-plane system sizes of $64\times 64$ sites. Fig.~\ref{fe2ir-sd-withb} shows the resulting spin configuration of the surface Fe layer for the simulation including biquadratic couplings. Note that due to the strong FM interlayer coupling 
the subsurface layer displays a very similar pattern. 

The Fourier transform of the spin configuration indicates that the propagation direction of the spin spirals is not exactly diagonal, since the wave vector is $\vec q=\left(0.13, 0.16\right)\!\frac{\pi}{a_\text{2d}}$ (and symmetry related points). The fact that $\vec q$ is not along any high symmetry lines is consistent with the degeneracy seen on the $J\!\left(\vec q\right)$ surface, however, the magnitude of the wave vector is twice
as large as that of the mean field estimate. This quantitative difference can mainly be attributed to the flaws of the mean field approach, but we can not rule out finite size effects in the spin dynamics simulation. 

The simulation without biquadratic interactions resulted in a very similar pattern indicating that these 
interactions are irrelevant in thicker films due to increased isotropic Heisenberg couplings. Simulations performed using only the isotropic part of the Heisenberg tensors led to the appearance of large FM domains without the spin spiral pattern seen in Fig.~\ref{fe2ir-sd-withb}. This result is corroborated by several previous
observations that the DM interaction may significantly affect the ground--state spin configuration, and even lead to the formation of spin spirals in thin film systems.\cite{ferriani-2008,udvardi-2008} It is also interesting
to note that a very similar labyrinth pattern was found in a monolayer of Mn on W(001).\cite{ferriani-2008} Composed of single-$q$ cycloidal spin spirals, the configuration shown in Fig.~\ref{fe2ir-sd-withb} does not have a net magnetic moment.

\subsection{Fe$_\mathbf{4}$/Ir(001)}

In the following, for the case of four monolayers of Fe, the Fe layers are going to be indexed 
from the substrate to the surface, so that layer 1 is closest to the Ir substrate and layer 4 is the surface layer. Summarizing the calculated interactions, the interlayer isotropic Heisenberg couplings are strong and FM, with increasing magnitude towards layers closer to the surface. Interestingly the couplings between intralayer first nearest neighbors are mostly weakly antiferromagnetic, only the two layers closest to the substrate show couplings of considerable magnitude, above 1~mRy in absolute value. The corrections to the Heisenberg spin-model are weak. Only the first NN\ intralayer DM vectors in layer 1 are larger than 0.1~mRy, and the strongest biquadratic couplings are around 0.07~mRy. Based on these features it is probable that the magnetic behavior is dominated by isotropic couplings.

\begin{figure}[ht]
\includegraphics[width=\linewidth]{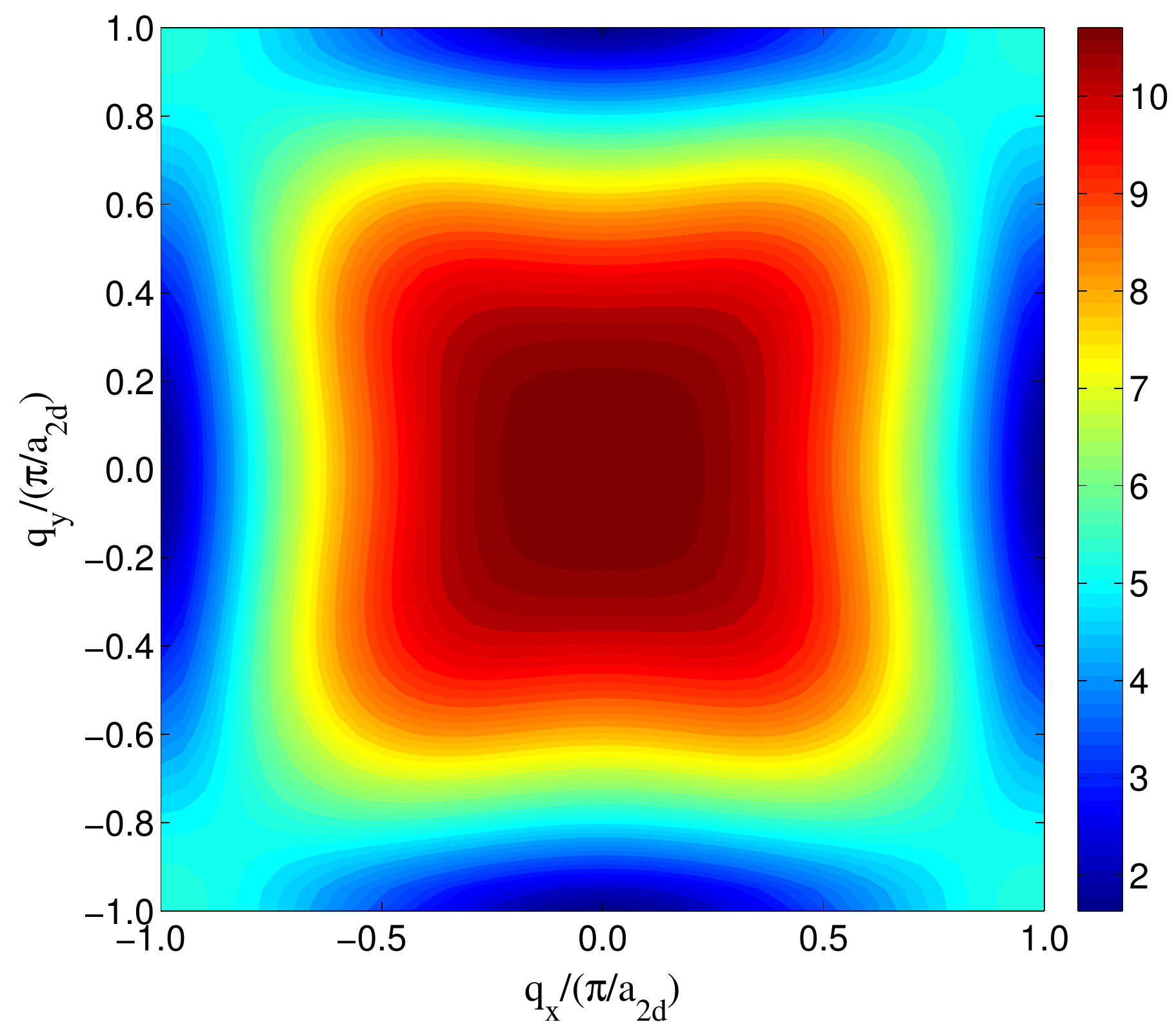}
\caption{(Color online) $J\!\left(\protect\vec q\right)$ surface for Fe$_4$/Ir(001). The color coding in units of mRy is shown on the right. }
\label{fe4ir-jq}
\end{figure}

The $J\!\left(\vec q\right)$ surface for Fe$_4$/Ir(001) shown in Fig.~\ref{fe4ir-jq} has its maximum in the $\Gamma$ point, anticipating FM arrangement. If the ground state is indeed FM, it is worth looking at the mean field estimate for the Curie temperature. The maximum of the surface with 10.89~mRy corresponds to a mean field estimate of $T_C=573$~K. As mean field approximations overestimate the stability of the ordered phases due to neglected fluctuations, the actual critical temperature is surely lower than this value. Still, it is possible that at room temperature the system is in the ordered phase.

The spin dynamics simulations revealed that the ground state is more complicated than what the mean field approach predicts. The simulated system converged into a complex spin structure regardless of the presence of biquadratic coupling terms. In each layer the spins formed a cycloidal spiral of $\vec q=\left(0.41, 0.41\right)\!\frac{\pi}{a_\text{2d}}$ rotating in a vertical plane, but with an additional in-plane FM modulation leading to nonzero net magnetization, $\left\langle \vec e_i\right\rangle\ne 0$. While the wave vector of the spiral is the same within every layer due to the strong interlayer coupling, the magnitude of the FM Fourier component increases with distance from the substrate. For a quantitative comparison of the net magnetizations, the layer-averaged magnetizations calculated from a homogeneous domain containing 1178 spins are presented in Table~\ref{fe4ir-sd-totmag}. The average magnetization, $\left\langle \vec e_i\right\rangle$, is clearly in-plane and monotonically increasing in magnitude towards the surface where it reaches a value larger than 0.7. Considering the actual size of each magnetic moment, also shown in Table~\ref{fe4ir-sd-totmag}, the total average magnetic moment of the system is 1.27$\mu_\text{B}$.

\begin{table}[htb]
\caption{Layer-resolved averaged magnetizations, $\left\langle \vec e_i\right\rangle$, 
and atomic magnetic moments in a given domain of Fe$_4$/Ir(001) containing 1178 spins.}
\label{fe4ir-sd-totmag}
\begin{ruledtabular}
\begin{tabular}{ccc}
& $\left\langle \vec e_i\right\rangle$ & magnetic moment [$\mu_\text{B}$]\\
\colrule
layer 1 & $\left(0.26,  0.26, 0.01\right)$ & 1.98\\
layer 2 & $\left(0.37,  0.37, 0.01\right)$ & 2.02\\
layer 3 & $\left(0.51,  0.53, 0.00\right)$ & 1.60\\
layer 4 & $\left(0.54,  0.57, 0.00\right)$ & 2.71\\
\end{tabular}
\end{ruledtabular}
\end{table}

In contrast to bulk bcc Fe, the net magnetization points along the $\left(110\right)$ direction, i.e., towards intralayer second-nearest neighbors, but it is indeed in-plane. Possibly in even thicker films we would see FM order with easy axis along the (100) direction, as was found experimentally.

\section{Conclusions}
We used the recently developed SCE-RDLM method to obtain spin Hamiltonians from \emph{ab initio} calculations, going beyond the anisotropic Heisenberg model by including isotropic biquadratic interaction terms. The obtained interaction parameters allow for a detailed investigation of the magnetic ground state via a mean field approach and atomistic spin dynamics simulations. 
We presented results for Fe thin films on the (001) surface of a semi-infinite Ir substrate. For the case of an Fe monolayer we found that layer relaxations drastically rearrange the interaction landscape, leading to the appearance of complex non-collinear spin structures. 
Relativistic corrections, in particular, Dzyaloshinskii-Moriya interactions, are needed to be taken 
into account as anisotropic couplings largely affect the ground state. Spin dynamics simulations also revealed that including biquadratic interactions to the spin model significantly alters the ground state spin configuration by favoring a more collinear state. This serves as an instructive warning that in some systems the usual Heisenberg model might give an insufficient description of magnetic properties.

For thicker films of two and four monolayers we found that biquadratic couplings are irrelevant in determining the ground--state spin configuration due to the large magnitude of the Heisenberg interaction. While the bilayer system still produces a single-$q$ spin spiral as ground state due to the DM interaction, in the quadrilayer system there is nonzero net magnetization superimposed on a cycloidal spin structure. This finding may be consistent with experimental results showing a ferromagnetic signal in MOKE measurements above four monolayers of Fe deposited on Ir(001). 

\begin{acknowledgments}
Financial support was provided by the Hungarian Research Foundation (contract no.\ OTKA K77771 and K84078) and by the New Sz\'echenyi Plan of Hungary (Project ID: T\'AMOP-4.2.1/B-09/1/KMR-2010-0002). Josef Kudrnovsk\'y and Vaclav Drchal are gratefully acknowledged for fruitful and stimulating discussions.
\end{acknowledgments}

\appendix*
\section{Mean field paramagnetic spin susceptibility}\label{app:spinsusc}
For a system of spins $\left\lbrace \vec e\right\rbrace$ governed by the grand potential 
$\Omega\left(\left\lbrace \vec e \right\rbrace\right)$, the best variational mean field trial Hamiltonian $\Omega_0\left(\left\lbrace \vec e \right\rbrace\right)=-\sum_i \vec h_i \vec e_i$, i.e., the one that minimizes the free energy is given by
\begin{align}
\vec h_i=-\frac{3}{4\pi}\int \mathrm d^2 e_i\; \vec e_i \left\langle \Omega_0 \right\rangle ^0 _{\vec e_i},
\end{align}
where the angle brackets now denote thermodynamic averaging with respect to the mean field probability distribution $P_0\!\left(\left\lbrace \vec e_i\right\rbrace \right)=\exp\left[-\beta \Omega_0\!\left(\left\lbrace \vec e_i\right\rbrace\right)\right]/Z_0$, $Z_0$ being the corresponding canonical partition function.

For our spin model extended with an inhomogeneous external field,
\begin{align}
\Omega\!\left(\left\lbrace \vec e\right\rbrace\right)&=\Omega_0 +\sum\limits_{i=1}^N \vec e_i \,\mathbf K_i \vec e_i -\frac{1}{2}\sum\limits_{\substack{i, j=1\\\left(i\ne j\right)}}^N \vec e_i \,\mathbf J_{ij}\vec e_j\notag\\
&\quad-\frac{1}{2}\sum\limits_{\substack{i, j=1\\\left(i\ne j\right)}} B_{ij}\!\left(\vec e_i \cdot \vec e_j\right)^2-\sum_{i=1}^N \vec h_i^\text{ext}\cdot \vec e_i,
\end{align}
the Weiss field reads as
\begin{align}
\vec h_i=\vec h_i^\text{MF}+\vec h_i^\text{ext},
\end{align}
where the molecular field induced by the interaction takes on the familiar mean field form,
\begin{align}
\vec h_i^\text{MF}=\sum\limits_{j\left(\ne i\right)} \mathbf J_{ij} \vec m_j,
\end{align}
$\vec m_j=\left\langle \vec e_j \right\rangle^0$ being the average magnetization at site $j$. Note that neither the on-site anisotropy nor the biquadratic couplings contribute to the Weiss field.

The spin susceptibility, defined as $\chi_{ij}^{\alpha\beta}=\partial m_i^\alpha/\partial h_j^\text{ext,$\beta$}$ may be related to the susceptibility of the non-interacting spin system $\chi_{ij}^{0,\alpha\beta}=\partial m_i^\alpha/\partial h_j^\beta$ as
\begin{align}
\bm \chi_{ij}=\bm\chi_{ij}^{0}+\sum\limits_{k\ne l}\bm \chi_{ik}^{0}\mathbf J_{kl}\bm \chi_{lj}.
\end{align}
For layered systems with only two-dimensional translation invariance we may separate site indices according to a layer index and an in-plane index,
\begin{align}
\bm \chi_{IJ,ij}=\bm\chi_{IJ,ij}^{0}+\sum\limits_{\substack{K,L,k,l\\(K,k)\ne (L,l)}}\bm \chi_{IK,ik}^{0}\,\mathbf J_{KL,kl}\bm \chi_{LJ,lj},\label{spinsusc-realspace}
\end{align}
where capital indices denote layers. Due to in-plane translation invariance we may introduce the two-dimensional Fourier transform of the quantities, for instance
\begin{align}
\bm \chi_{IJ} \!\left(\vec q\right)=\sum\limits_{\vec R_i} \bm \chi_{IJ,i0} \operatorname{e}^{-\imath \vec q\vec R_i},
\end{align}
which might be thought of as blocks of matrices in layer and Cartesian space,
\begin{align}
\left[\widehat{\bm \chi} \!\left(\vec q\right)\right]_{IJ}=\bm \chi_{IJ} \!\left(\vec q\right).
\end{align}

Using this notation we may easily invert Eq.~\eqref{spinsusc-realspace} and arrive at
\begin{align}
\widehat{\bm \chi}\!\left(\vec q\right)&=\left(\widehat{\mathbf I}- \widehat{\bm \chi}^{0}\!\left(\vec q\right) \widehat{\mathbf J}\!\left(\vec q\right)\right)^{-1} \widehat{\bm \chi}^0\!\left(\vec q\right),
\end{align}
indicating an enhancement of the spin susceptibility due to the Heisenberg interaction. In the paramagnetic limit this simplifies to
\begin{align}
\widehat{\bm \chi}\!\left(\vec q\right)&=\left(3 k_B T\widehat{\mathbf I}- \widehat{\mathbf J}\!\left(\vec q\right)\right)^{-1}.
\end{align}
Annealing the system from the paramagnetic phase there will be a temperature,
$T_\text{ord}$, where $\widehat{\bm \chi}\!\left(\vec q\right)$ becomes singular, indicating that the paramagnetic phase is unstable to the formation of some magnetically ordered state at that temperature. This transition temperature $T_\text{ord}$ is given by the condition
\begin{align}
T_\text{ord}&=\frac{1}{3 k_B} \max\limits_{\vec q} \left\lbrace \text{eigenvalues of $\widehat{\mathbf J}\!\left(\vec q\right)$}\right\rbrace\notag\\
&=\frac{1}{3 k_B} \max\limits_{\vec q} J\!\left(\vec q\right),\label{MF-est}
\end{align}
and the corresponding $\vec q$ where this happens gives the characteristic wave vector of the ordered phase (where the $\vec q$ vectors are sought for in the two-dimensional BZ). In Eq.~\eqref{MF-est} we introduced the notation $J\!\left(\vec q\right)$ for the maximal eigenvalue of $\widehat{\mathbf J}\!\left(\vec q\right)$ at a given wave vector. We may gain valuable insight regarding magnetic ordering in the system by plotting this scalar function against the points of the BZ.

\end{document}